\documentclass[aps,prb,twocolumn,groupedaddress,showpacs,notitlepage]{revtex4-1}
\usepackage{hyperref}
\usepackage{amsmath}
\usepackage{graphicx}

\usepackage[T1]{fontenc}
\usepackage[latin1]{inputenc}

\newcommand{\be}{\begin{equation}}
\newcommand{\ee}{\end{equation}}

\begin{document}

\title{Flat band states: disorder and nonlinearity}

\author{Daniel Leykam$^1$, Sergej Flach$^2$, Omri Bahat-Treidel$^3$, and Anton S. Desyatnikov$^1$}

\affiliation{$^1$Nonlinear Physics Centre, Research School of Physics and Engineering, The Australian National University, Canberra ACT 0200, Australia
\\
$^2$New Zealand Institute for Advanced Study, Centre for Theoretical Chemistry and Physics, Massey University, Auckland 0745, New Zealand
\\
$^3$School of Mathematics and Physics, The University of Queensland, Brisbane QLD 4072, Australia}
\date{\today}

\begin{abstract}
We study the critical behaviour of Anderson localized modes near intersecting flat and dispersive bands in the quasi-one-dimensional diamond ladder with weak diagonal disorder $W$. The localization length $\xi$ of the flat band states scales with disorder as $\xi \sim W^{-\gamma}$, with $\gamma \approx 1.3$, in contrast to the dispersive bands with $\gamma =2$. A small fraction of dispersive modes mixed with the flat band states is responsible for the unusual scaling. Anderson localization is therefore controlled by two different length scales. Nonlinearity can produce qualitatively different wave spreading regimes, from enhanced expansion to resonant tunneling and self-trapping.
\end{abstract}


\maketitle

\section{Introduction}

Disorder suppresses propagation of waves, resulting in the celebrated Anderson localization~\cite{anderson, kramer1993}. Nonlinearity has a profound effect on Anderson localized modes, creating chaos and delocalization \cite{pikovsky2008, flach2009, michaely2012}, self-trapping~\cite{aubry2008, conti2012}, or a combination of the two~\cite{aubry2008}. The mechanisms and details of these processes remain inconclusive and the contradictions are hotly debated~\cite{fishman2012}. Understanding the competition between disorder and interactions promises a wealth of applications because in any realistic system both are always present.

Unconventional Anderson localization is expected in systems containing dispersionless or {\em flat} bands~\cite{derzhko2006,derzhko2010,nita2013}. Weak disorder lifts the degeneracy, and accounts for both hybridization and disorder in a non-perturbative way. Energetically isolated flat bands result in an ``inverse'' Anderson transition in three dimensions (3D), where hybridization wins, and localized flat band states (FBS) delocalize with increasing disorder~\cite{goda2006, nishino2007}. A very specific situation arises when the flat band touches other dispersive bands at a point of zero group velocity \cite{chalker2010}. Numerical calculations in a 2D lattice in the limit of weak disorder revealed critical, multifractal FBS, reminiscent of an Anderson transition. This is quite different from ordinary 1D and 2D lattices, which require long range coupling for critical behaviour to appear \cite{mirlin_review}.

These results highlight the unusual consequences of mixing macroscopically degenerate FBS via disorder. They also show that the mixing is sensitive to both the dimensionality of the system and the inclusion of a small number of modes which belong to dispersive bands. So far the most interesting case of FBS fully immersed in a dispersive band structure has not been studied. Rigorous analytic results are scarce, and numerical studies in two or more dimensions are notoriously hard and imprecise due to finite size effects. Also, all studies of such systems to date have been limited to linear waves. How do nonlinearity or interactions affect a disordered flat band?

In this paper, we study wave localization and transport in a quasi-1D system, the diamond ladder, which hosts intersecting flat and dispersive bands. We show how disorder-induced mixing between flat and dispersive band states (DBS) produces Cauchy distributed disorder, heavy tailed statistics, multiple localization length scales and sparse, multi-peaked modes in the weak disorder limit. This has profound effects on wavepacket spreading in the presence of nonlinearities. A huge advantage compared to higher dimensional lattices is that here we obtain rigorous numerical results, free of finite size effects. This relatively simple lattice model involving only short-range couplings can be readily implemented in a variety of systems, such as optical waveguide arrays~\cite{christodoulides2003,segev2007,lahini2008,nature_review}, microwave resonators~\cite{bellec13}, exciton-polariton condensates \cite{masumoto2012}, and optical lattices for ultracold atomic gases~\cite{bloch2008,hyrkas2013}.

Our main finding is that even weak mixing between the dispersive and flat bands completely changes the transport properties of the system. The effective disorder potential for FBS has heavy Cauchy tails and correlations. The localization length $\xi$ at the flat band centre scales with disorder $W$ as $\xi \sim W^{-\gamma}$ with exponent $\gamma \approx 1.3$, while dispersive modes yield the usual $\gamma=2$ exponent \cite{kramer1993}. Therefore, the localization is governed by different length scales. Flat band modes are highly sparse, with multiple peaks, resulting in strong fluctuations in transport properties. Introducing a gap $m > W$ suppresses mixing with the dispersive bands, and we find that FBS no longer scale with disorder, $\gamma =0$, with compact profiles and small fluctuations resembling ordinary Anderson localization. On the other hand, the nonlinear mixing has an \emph{opposite} effect: mixing between FBS enhances fluctuations, while mixing with DBS reduces them. Thus, qualitatively different wave spreading regimes are tunable via the interaction strength.

Sec. \ref{sec-model} introduces our model and examines the properties of the linear modes of the disordered system. In Sec. \ref{sec-spreading} we explore the spreading of localized excitations as a function of the nonlinearity strength. Sec. \ref{sec-final} concludes the paper with discussion of future directions and possible experimental realizations of our model, and a summary of our results. 

\section{Model \& linear modes}
\label{sec-model}
\begin{figure}
\includegraphics[width=\columnwidth]{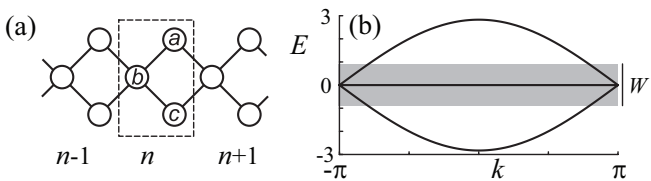}
\caption{(a) Lattice structure with $a$,$b$,$c$ sublattices, unit cell marked by dashed line. (b) Band structure. Disorder $W$ smears out the flat band to a width $W$ (shaded region).}
\label{lattice}
\end{figure}

The diamond ladder is shown in Fig.~\ref{lattice}(a). Propagating waves can travel along two possible paths, through either the ``a'' or ``c'' sites. Destructive interference between these two paths can be introduced in a variety of ways, for example by applying a magnetic field \cite{vidal2000} or Rashba spin-orbit coupling \cite{bercioux2004}, which results in wave localization. Interesting interacting phases have also been obtained in the corresponding Hubbard \cite{gulacsi2007,lopes2011, rojas2012,hyrkas2013} and Ising models \cite{schmidt2013}. Here, we consider a tight binding model with mean field interaction terms, which hosts intersecting dispersive and flat bands at the Brillouin zone edge [Fig.~\ref{lattice}(b)],
\begin{align}
&i \dot{\bf a}_n + (\epsilon_{a,n}+\beta{\bf |a}_n|^2){\bf a}_n = - \nabla^2   {\bf b}_{n+1} , \label{m1-1}\\
&i \dot{\bf b}_n + (\epsilon_{b,n}+\beta |{\bf b}_n|^2){\bf b}_n =  - \nabla^2({\bf a}_n + {\bf c}_n) , \label{m1-2} \\
&i \dot{\bf c}_n + (\epsilon_{c,n}+\beta|{\bf c}_n|^2){\bf c}_n =  - \nabla^2 {\bf b}_{n+1} , \label{m1-3}
\end{align}
here $\nabla^2 f_n = f_n + f_{n-1}$ is the discrete Laplacian, $\beta$ is the nonlinearity coefficient, and $\epsilon_{j,n}$ is the disorder potential, $j=a,b,c$. The dot denotes derivative with respect to time $t$ or propagation length in optical waveguide arrays. We set the conserved norm $\sum_n (|{\bf a}_n|^2 + |{\bf b}_n|^2 + |{\bf c}_n|^2)$ to 1 without loss of generality. We can also choose $\beta > 0$ (attractive nonlinearity), as equivalent results for $\beta < 0$ may be obtained by applying the staggering transform $\beta \rightarrow -\beta, \epsilon_n \rightarrow -\epsilon_n, b_n \rightarrow -b_n$. 

In the linear, disorder-free limit $\beta=\epsilon_{j,n}=0$, the mode profiles $\psi_n = \{a_n,b_n,c_n\}$ are found from Eqs.~\eqref{m1-1}-\eqref{m1-3} using $\{{\bf a}_n(t),{\bf b}_n(t),{\bf c}_n(t)\}= \psi_n e^{i E t}$. The linear spectrum $E(k)= 0, \pm 2 \sqrt{2} \cos (k / 2)$ in Fig.~\ref{lattice}(b) is derived using plane wave expansion $\psi_n=\psi e^{ikn}$ with wavenumber $k$. FBS take the form 
\be 
\psi_n = \{1,0,-1\} f_n, \label{fb}
\ee
where $f_n$ is an arbitrary function. For example, $f_n = \delta_{n,n_0}$ gives a compact mode perfectly localized to a single unit cell $n_0$. The $\pi$ phase difference between the ``a'' and ``c'' sites causes destructive interference which effectively decouples sublattices and prevents diffraction. DBS with $E\ne 0$ are infinitely extended, $\psi_n=e^{ikn}\{1,\pm\sqrt{2}e^{-ik/2},1\}/2$. We note that here the flat band is robust against direct coupling between the ``a'' and ``c'' sublattices. The most important requirement is that the next-nearest neighbour coupling (eg. between ``a'' sublattice sites) should be small.

We consider diagonal disorder with uncorrelated, uniformly distributed random variables $\epsilon_{j,n} \in [-W/2,W/2]$. The spectrum is bounded by $[-\Delta,\Delta]$, with spectral width $\Delta = 2 \sqrt{2} + W/2$.

It is convenient to solve Eqs.~\eqref{m1-1},~\eqref{m1-3},
\be 
a_n = \frac{b_n + b_{n+1}}{E - \epsilon_{a,n}}, \qquad c_n = \frac{b_n + b_{n+1}}{E - \epsilon_{c,n}},
\ee
to obtain a single equation for a mode profile $b_n$:
\begin{align}
& \varepsilon_n b_n = C_n b_{n+1} + C_{n-1} b_{n-1}  \label{d1} \\
& C_n = (\epsilon_{a,n} - E)^{-1} + (\epsilon_{c,n} - E)^{-1}, \label{d2} \\
& \varepsilon_n = \epsilon_{b,n} - E - C_n - C_{n-1}, \label{d3}
\end{align}
which, at first glance, resembles an ordinary periodic 1D lattice with both diagonal, $\varepsilon_n$, and coupling, $C_n$, disorder~\cite{kramer1993}. In addition, there are short range correlations between the two.

This effective disorder acquires specific structure with two distinct energy regimes. For low energy $|E| < W/2$, the couplings $C_n$ can vanish or diverge, resulting in non-perturbative behaviour. The probability distribution function (PDF) in this case, $f(y)$ of $y=(\epsilon_{j,n} - E)^{-1}$, is nonzero only at the tails, $|1/y+E| \leq W/2$, where it is given by $f(y)=1/(Wy^2)$. It is similar to the Cauchy distribution for large arguments, with Cauchy-like ``heavy tails'' decaying slowly as $1/y^2$, such that its variance diverges. The PDFs for $C_n$ and $\varepsilon_n$ also have this heavy tail.  Modes in this energy range are primarily composed of FBS and the non-perturbative behaviour is due to their macroscopic degeneracy when $W=0$. In contrast, the dispersive bands provide the dominant contribution for high energy $|E| > W/2$, with all couplings $C_n$ finite, and all relevant PDFs lacking the above Cauchy-like tails.

Numerically, we diagonalize Eqs.~\eqref{m1-1}-\eqref{m1-3} for a disordered chain of finite size $N$ with periodic boundary conditions \cite{details} and obtain the modes $\psi_{\nu,n} = \{a_{\nu, n},b_{\nu,n}, c_{\nu,n}\}$, $\nu=1, 2\dots 3N$. We characterize mode behaviour as a function of $E$ using the following measures~\cite{krimer2010}: The participation ratio
\be 
P = 1/ \sum_n ( |a_{\nu,n}|^4 + |b_{\nu,n}|^4 + |c_{\nu,n}|^4),
\ee
measures the number of strongly excited sites. The second moment
\be 
m_2 = \sum_n [(X_{\nu} - n)^2 |b_{\nu,n}|^2 + (X_{\nu} - n - 1/2)^2 (|a_{\nu,n}|^2 + |c_{\nu,n}|^2)],
\ee
is sensitive to the distance between the tails of the eigenmode, here $X_{\nu} = \sum_n [n |b_{\nu,n}|^2 + (n + 1/2) ( |a_{\nu,n}|^2 + |c_{\nu,n}|^2)]$ is the mode's centre of mass. The compactness index
\be 
\zeta = P^2 / m_2,
\ee
reveals how uniformly the eigenstate excites the volume it occupies. We calculate the mean values of $P$, $m_2$, and $\zeta$ for each value of $W$ by taking a sample of $\sim 100,000$ modes divided into 100 energy bins.

We also obtain the localization length $\xi$ (asymptotic decay rate of the eigenmode tails, $\psi_{\nu, n} \sim e^{-n / \xi}$) by applying
\be
\xi^{-1} ( E ) = \lim_{N \rightarrow \infty} \frac{1}{N} \left\langle \sum_{n=1}^N \ln \tfrac{b_{n+1}}{b_n}  \right\rangle,
\ee
where $\langle \cdot \rangle$ denotes averaging over different realizations of disorder\cite{average}.

\begin{figure}
\includegraphics[width=1 \columnwidth]{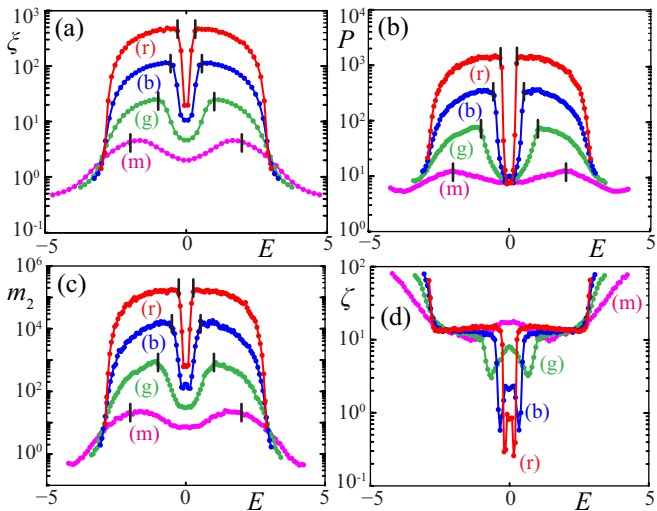}
\caption{(color online) Mode properties: (a) mean localization length $\xi$, (b) participation ratio $P$, (c) second moment $m_2$, and (d) compactness index $\zeta$ as a function of energy $E$ for different disorder strengths: $W$ = 0.5 [(r)ed], 1 [(b)lue], 2 [(g)reen], 4 [(m)agenta]. Statistical errors do not exceed the spot size. Vertical bars indicate the cutoff energy $|E| = W/2$.}
\label{loc_length}
\end{figure}

Figure~\ref{loc_length} shows the results for different disorder strengths. Indeed, the boundary $|E| = W/2$ (marked by vertical bars) separates modes with very different properties. For high energy, $|E| > W/2$, we obtain compact, weakly localized modes with properties similar to those of an ordinary weakly disordered 1D chain.

\begin{figure}
\includegraphics[width= \columnwidth]{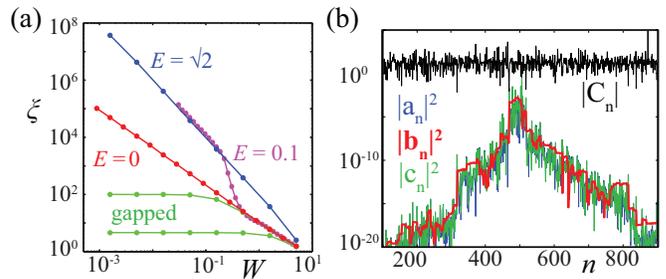}
\caption{(color online) (a) Localization length $\xi (W)$ for low ($E=0$) and high ($E=\sqrt{2}$) energy modes, following the power laws $\xi \sim W^{-1.3}$ and $\xi \sim W^{-2}$ respectively. The $E=0.1$ curve shows the crossover between these laws at $W \sim |E|$. $\xi ( E = 0 )$ remains finite in the gapped cases $m = 0.5$ (lower green curve) and $m=0.05$ (upper green curve). (b) Mode profile of a sparse ($\zeta = 0.1$) low energy mode, $W=0.5$, and the effective coupling $|C_n|$ defined by Eq.~(6).}
\label{fig3}
\end{figure}

When the energy is low, $|E| < W/2$, the modes display remarkably different properties: $\xi$, $P$ and $m_2$ are orders of magnitude smaller, suggesting much stronger localization. However, the compactness index $\zeta$ is also very small, indicating sparse modes consisting of well-separated peaks, completely different from conventional Anderson localized modes. We obtain the power law $\xi (W) \sim W^{-\gamma}$ [Fig.~\ref{fig3}(a)], with surprising scaling $\gamma = 1.30 \pm 0.01$ at $E=0$, in contrast to the usual $\gamma = 2$ in the dispersive bands. We further note that this is clearly different from the value $\gamma=1$ which is expected for a 1D chain with Cauchy-distributed onsite energies~\cite{lloyd1969,thouless1972,ishii1973}. If we move slightly away from the flat band energy, $E \ne 0$, we observe this anomalous exponent as long as $W \gtrsim |E|$, ie. $E$ lies within the disorder-broadened flat band. As $W$ is decreased further, there is a rapid crossover to the conventional $\gamma = 2$ scaling. The other measures $P$, $m_2$, and $\zeta$ similarly display anomalous scaling at $E=0$: for example, $P\approx 7$ does not change at all with $W$. Thus, some measures suggest localized modes (finite $P$), while others suggest extended states ($m_2$ diverges). Hence the FBS display criticality in the weak disorder limit.

Plotting the typical profile of a low energy mode in Fig.~\ref{fig3}(b), we observe exponential localization combined with strong fluctuations in the amplitudes on the ``a'' and ``c'' sublattices. These strong fluctuations are a signature of the heavy-tailed effective disorder \cite{titov}, and occur when the coupling $C_n$ in Eq.~\eqref{d2} is small. Between these points, $b_n$ stays remarkably constant. Thus, one can view the low energy modes as combinations of highly localized flat band components (the ``a'',``c'' spikes) whose coupling together is mediated by a small, weakly localized dispersive band component.

This coupling is frozen if the flat and dispersive bands are separated by a gap, because the dispersive states near $E=0$ become strongly localized \cite{flach_arxiv}. We create a gap by introducing equal and opposite mass terms at the ``a'' and ``c'' sublattices by shifting the onsite potentials, $\epsilon_{a,n} \to \epsilon_{a,n} + m$ and $\epsilon_{c,n} \to \epsilon_{c,n} - m$. The modified spectrum at $W=0$ is $E(k) = 0, \pm \sqrt{m^2 + 1 + \cos k }$, which preserves the flat band, while creating gaps of size $m$ with the dispersive bands. We plot $\xi (W)$ with a mass term $m = 0.5$ and $m=0.05$ at energy $E=0$ in Fig.~\ref{fig3}(a). For $W \ll m$, $\xi$ converges to a constant value. The modes freeze their localization length and resemble normal Anderson modes. The localization length at $W\rightarrow 0$ depends on the gap size $m$, which controls the localization of dispersive states at $E=0$. $P$, $m_2$ and $\zeta$ are also converge to constants. Thus at small $W$ the random potential introduces hybridization and disorder in a balanced way for gapped FBS. In return, in the absence of a gap the mixing with the DBS takes over the effective disorder and is responsible for the critical behaviour of the low energy modes.

\section{Dynamics}
\label{sec-spreading}

\begin{figure}
\includegraphics[width= \columnwidth]{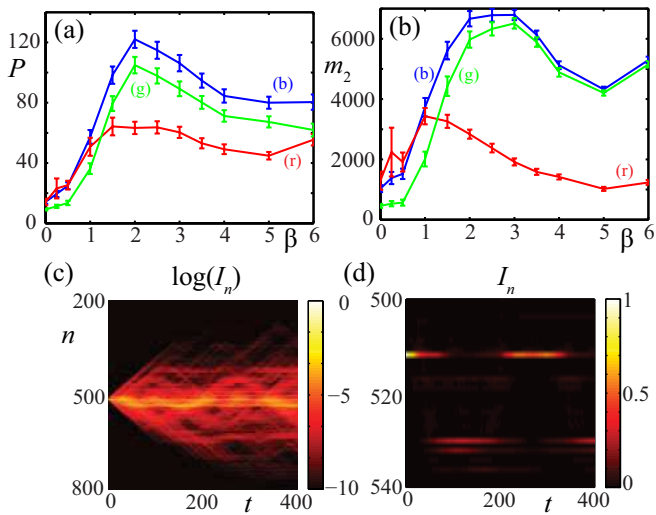}
\caption{(color online) (a) Participation ratio of a flat band excitation after evolution for $t=400$ as a function of nonlinearity strength, $\beta$. For each point, we obtain 500 realizations of disorder, and show the mean $\langle P \rangle$ [(b)lue], standard deviation [(r)ed], and typical value $\exp ( \langle \ln P \rangle )$ [(g)reen]. Error bars indicate 95\% confidence intervals obtained via bootstrap method. (b) Same, but second moment instead. (c) Intensity $I_n = |a_n|^2 + |b_n|^2 + |c_n|^2$ during linear propagation when sparse modes are strongly excited (d) Same, but linear scale. In all panels, $W=1$.}
\label{fig4a}
\end{figure}

The critical mode properties are essential in understanding the linear and nonlinear transport properties of the system. In the following, we consider the expansion of a flat band excitation [Eq. \eqref{fb}] initially localized to a single unit cell in a moderately disordered ($W=1$) system. We quantify the spreading by calculating $P$ and $m_2$ at $t=400$, which is long enough for initial transients to die out. Repeating for many realizations of disorder, we obtain distributions, which we characterize via their mean, standard deviation and typical values. Results as a function of nonlinearity strength $\beta$ are plotted in Fig. \ref{fig4a}(a,b).

With weak nonlinearity, $P$ and $m_2$ present heavy-tailed distributions: the mean values significantly exceed the typical values and are comparable to the standard deviation. In other words, the fluctuations in the wavepacket spreading are large. Hence, even with moderate disorder strength, a strong signature of the critical behaviour in the weak disorder limit persists. Furthermore, weak nonlinearity tends to \emph{amplify} the heavy-tailed nature of the spreading, increasing the mean more than the typical value.

To understand this behaviour, it is instructive to study the dynamics of individual realizations. Most of the time, an ordinary, diffusive expansion to a size on the order of a localization length occurs. However, if highly sparse modes such as the one shown in Fig.~\ref{fig3}(b) are strongly excited, much larger expansion driven by tunnelling to distant sites occurs, including the oscillation of energy back and forth between well separated peaks in Figs.~\ref{fig4a}(c,d). In contrast to conventional 1D lattices \cite{veksler2010}, these sparse modes are statistically significant. This is the origin of the heavy tail in the spreading behavior.

Above a threshold interaction strength $\beta \approx W/2$, the expansion grows significantly, and typical values of $P, m_2$ approach their means, indicating the emergence of more normal statistics. Thus, weak nonlinearity enhances the critical behaviour, but stronger nonlinearity suppresses it. Increasing $\beta$ further still, the expansion reaches a maximum, then starts to decrease.

To explain these different regimes of nonlinearity, we consider Eqs.~\eqref{m1-1}-\eqref{m1-3} in the basis of eigenmodes of the linear system $\psi_{\nu,n}$~\cite{krimer2010},
\be
i \dot{\phi}_{\nu} = E_{\nu} \phi_{\nu} + \beta \sum_{\nu_1,\nu_2,\nu_3} I_{\nu,\nu_1,\nu_2,\nu_3} \phi_{\nu_1}^* \phi_{\nu_2} \phi_{\nu_3}, \label{n1}
\ee
where $\phi_{\nu}(t)$ is the complex amplitude of mode $\nu$ and $I$ is the overlap integral
\be
I_{\nu,\nu_1,\nu_2,\nu_3} = \sum_n \sum_{\alpha = a,b,c} \alpha_{\nu,n}^* \alpha_{\nu_1,n}^* \alpha_{\nu_2,n} \alpha_{\nu_3,n}, \label{n2}
\ee
with the summation over all unit cells and sublattices. $I$ determines the effective strength of coupling between different modes~\cite{krimer2010}. In the disorder-free limit, $I$ can be calculated explicitly for a chain of size $N$. The coupling between dispersive band modes is subject to the selection rule $k^{\prime} + k_1 - k_2 - k_3 = 2 \pi n$, where $n$ is an integer, while the overlap between flat band states vanishes because they all occupy different lattice sites. Furthermore, the coupling between flat band and dispersive states also vanishes unless some dispersive band states are already excited. Therefore a pure excitation of the flat band will not spread at all, even in the presence of nonlinearity.

When disorder is introduced, these selection rules are broken, so the coupling can become stronger. Nonlinearity then introduces an additional energy scale that competes with the disorder, an energy shift $\delta E_{\nu} \approx s \beta I_{\nu,\nu,\nu,\nu} \approx s \beta / \sqrt{P}$, where $s = |\phi_{\nu}|^2$ is the occupation a given mode. High energy modes have diverging $P$ in the limit of weak disorder, leading to small shifts. On the other hand, the low energy modes with $P \sim 7$ can experience significant energy shifts.

With weak nonlinearity, $s \beta < W/2$, the nonlinear frequency shift does not exceed the width of the low energy subspace. Strong resonant interactions can only occur between low energy modes. The expansion due to resonant tunnelling shown in Fig. \ref{fig4a}(c,d) can either be enhanced or suppressed \cite{veksler2010}. Thus, fluctuations become more pronounced.

In the intermediate regime, $s \beta > W/2$, the nonlinear energy shift exceeds the width of the low energy subspace. Strong resonant energy transfer from low to high energy modes is responsible for the enhanced spreading and growth of $P, m_2$: the high energy modes are not strongly localized. Additionally, each flat band mode can transfer energy into \emph{many} dispersive band modes. Thus, a kind of self-averaging can occur, which is responsible for the more normal spreading statistics. Since $s$ decreases, at some point it will tune out of strong interaction with the dispersive band, leaving a flat band component which remains strongly localized for potentially long times.

For very strong nonlinearity the energy shift $\delta E $ can exceed the total band width, leading to self-trapping. Thus, $P$ and $m_2$ reach a maximum and then start to decrease.

\begin{figure}
\includegraphics[width=\columnwidth]{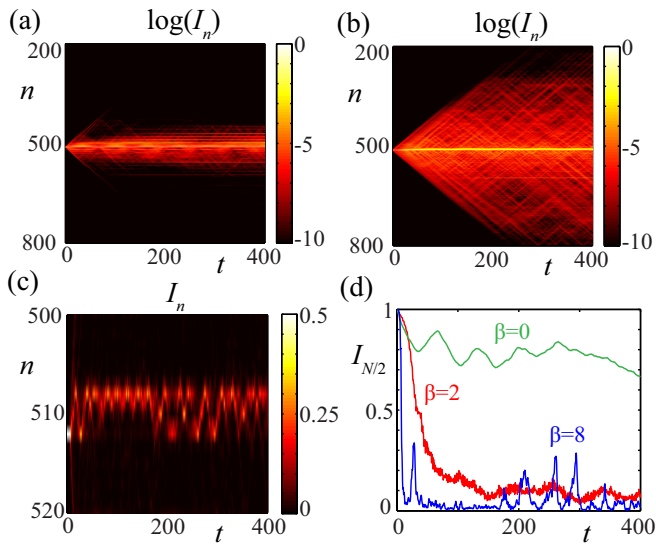}
\caption{(color online) Effect of nonlinearity on spreading of a single site excitation, $W=1$. (a) Linear propagation, $\beta = 0$. (b) Intermediate nonlinearity, $\beta = 2$. (c) Strong nonlinearity, $\beta = 8$. (d) The intensity at the initially excited unit cell in all three cases.}
\label{nonlinear}
\end{figure}

We illustrate these different regimes by presenting examples of propagation in Fig.~\ref{nonlinear}. Here, a single-peaked FBS is strongly excited. Under weak nonlinearity there is no resonant interaction with the dispersive bands and the wavepacket expansion is similar to the linear case in Fig.~\ref{nonlinear}(a). In the intermediate regime in Fig.~\ref{nonlinear}(b), the expansion is driven by an initial transfer of energy to the dispersive bands and we see stronger spreading. With strong nonlinearity in Fig.~\ref{nonlinear}(c) we observe the formation of a self-trapped state, which irregularly meanders between two quasi-stable positions. Energy is lost during this motion, and it eventually becomes trapped at a ``b'' sublattice site. We plot the intensity at the initially excited cell for these three cases in Fig.~\ref{nonlinear}(d) - observe how nonlinearity leads to a rapid transfer of energy away, leaving behind a small self-trapped component which persists for long times.

\section{Discussion \& conclusions}
\label{sec-final}

Flat band systems such as the diamond ladder are attracting growing interest as a means of realizing exotic strongly interacting phases of matter \cite{bergholtz13}. While we have considered in this paper a relatively simple tight binding model, we have verified in a number of other quasi-1D cases \cite{flach_arxiv} that our results should be generic to any system with intersecting flat and dispersive bands. Similarly, the emergence of different dynamical regimes due to the competition between disorder and nonlinearity should be a generic feature of other types of interaction terms, so it would be interesting to extend recent results on interacting bosons and fermions in the ideal diamond chain \cite{vidal2000,lopes2011,rojas2012,hyrkas2013} to disordered systems.

There are a variety of settings in which this type of tight binding model may be realized. Recently structured etching of microcavities has been used to fabricate 2D kagome lattice structures with a flat band for exciton-polariton condensates \cite{masumoto2012}. The same technique can also be applied to generate quasi-1D lattices such as the diamond ladder. Another approach is to use optical waveguide arrays, where a 1D flat band could be introduced by generalizing a single bound state in the continuum \cite{bic} to a large collection of degenerate bound states, and 2D flat band lattices (eg. kagome\cite{kagome} or Lieb\cite{leykam2012}) are also accessible. Similar techniques can be applied using optical traps for cold atoms \cite{apaja2010,hyrkas2013} and microwave resonator lattices \cite{bellec13}. Flat band-induced localization in 2D was also studied \cite{vidal2001} and observed in magnetic field-induced Aharonov-Bohm cages in superconducting wire networks \cite{ab-1} and AlGaAs-GaAs heterojunctions \cite{ab-2}. In this case, applying a magnetic field provides an alternative way to introduce a gap between the flat and dispersive bands.

To summarize, the diamond ladder presents a test bed for exploring the interplay between macroscopic degeneracy, disorder and nonlinearity. We showed how the mixing between macroscopically degenerate flat band modes and a small number of weakly localized modes of intersecting dispersive bands results in low energy modes with highly unusual properties. Consequently, the spreading of low energy wavepackets becomes sensitive to nonlinearity or interactions. Therefore, our results provide novel ideas for future studies in higher lattice dimensions and they highlight the importance of nonlinear interactions and many body quantum dynamics for weakly disordered flat band systems.

This work was supported by the Australian Research Council.


\begin{thebibliography}{99}
\bibitem{average}
The disorder average is not essential; we checked that this expression is self-averaging in the limit of large $N$.
\end{thebibliography}
\end{document}